\documentclass[twocolumn]{revtex4-1}
\usepackage{graphics}  
\usepackage{graphicx}  %
\usepackage{amsfonts}  
\usepackage{epsfig}    
\usepackage{color}     
\usepackage{amsmath}
\usepackage{mathtools}
\usepackage{subfigure}
\usepackage{verbatim}
\begin{document}

\title{Evaluation of a measure on the quasi-steady state assumption of 
Collisional Radiative Models via Intrinsic Low Dimensional Manifold 
Technique}
\author{Efe Kemaneci, Emile Carbone, Wouter Graef, 
Jan van Dijk, Gerrit M W Kroesen}

\affiliation{\small Department of Applied Physics, Eindhoven University of            
Technology, The Netherlands }                   
\date{\today}

\begin{abstract}

\noindent Collisional and radiative dynamics of a plasma is exposed 
by so-called Collisional Radiative Models \cite{Dij2001/2} that 
simplify the chemical kinetics by quasi-steady state assignment 
on certain types of particles. The assignment is conventionally 
based on the classification of the plasma species by the ratio 
of the transport to the local destruction frequencies. We show 
that the classification is not exact due to the role of the 
time-dependent local production, and a measure is necessary to 
confirm the validity of the assignment. The main goal of this 
study is to evaluate a measure on the quasi-steady state 
assumptions of these models. Inspired by a chemical reduction 
technique called Intrinsic Low Dimensional Manifolds 
\cite{Egg95,Maa92}, an estimate local source is provided at the
transport time-scale. This source is a deviation from the 
quasi-steady state for the particle and its value is assigned 
as an error of the quasi-steady state assumption. The propagation 
of this error on the derived quantities is formulated in the 
Collisional Radiative Models. Based on the error a novel technique 
is proposed to discriminate the quasi-steady states. The developed 
analysis is applied to mercury and argon fluorescent lamps 
separately and the corresponding errors are presented. We observe 
that the novel and conventional technique agrees for most of the 
excited levels but disagrees for a few low energy excited states. 

\end{abstract}

\maketitle

\section{Introduction}

Plasmas are extensively employed for the purpose of surface treatment 
\cite{Lie2005, Kup80} and lighting applications \cite{Fle06}. Their 
collective feature is a large number of locally interacting species 
such as excited atomic or molecular states and chemical reactants. 
This characteristic induces numerical difficulties such as heavy 
computational load and stiffness; therefore, long-lasting simulations 
and troublesome convergences appear. Although models that contain 
large number of species are still viable, these difficulties persuade 
reduction of this number. 

A naive approach in reducing the number of species is to ignore 
low-valued particle densities. Although it decreases the 
computational load and relieve the stiffness, it may lock essential 
reaction channels in the physical system. Hence, reduction 
techniques respecting all the local interactions are obligatory.   
{\em Collisional Radiative Model} (CRM) is such a reduction method, 
which is widely used in low-temperature atomic plasmas. It focuses 
on the interplay between collisional and radiative processes. The 
technique was commenced by Bates {\em et al.} \cite{Bat62} and 
improved further by various authors \cite{Mul90/3, Ben91, Dij2001/1}. 
So-called {\em Local Chemistry} (LC) species already equilibrated  
to {\em Quasi-Steady State} (QSS) in local chemistry at the transport
time-scale and it uses this fact to determine a reduced set of 
species and effective reaction rates. 

Reduction methods are utilised not only in the area of plasma science 
but also in other fields with overlapping peculiarity 
\cite{Maa92, Sti98, Kap2002}.
Though developed independently, some of these techniques are 
remarkably similar. For example, in mathematical biology, some
\cite{Sti98} use QSS assumptions, while certain combustion models 
\cite{Pet93} similarly assume partial equilibrium and steady state. 
Another reduction method that is extensively employed in combustion 
engine flame models is {\em Intrinsic Low Dimensional Manifold} 
(ILDM) technique. It was initiated by Maas {\em et al.} \cite{Maa92} 
and studied further by numerous authors \cite{Egg95, Sko2001}. It 
makes use of characteristic principal frequencies of local interactions 
to determine an asymptotic state of the mass fractions that leads 
the chemical reduction.

CRMs are strong tools with a low computational load for the 
investigation of the plasma species and their interactions, however,
they suffer from a couple of inadequacies. The first one is the LC 
condition, which discriminates the species quickly reaching QSS. It 
is based on the ratio of the transport to local destruction 
frequencies. Since it excludes the role of the time-dependent local 
production, it is insufficient in the assignments. The second 
inadequacy is the deviations from the QSS assumption that perturbs 
the derived quantities in the CRM. These deviations and the resulting 
perturbations are ignored in the conventional CRM formalism. In this 
study, these inadequacies are addressed with the aid of the ILDM 
technique.

Inspired by the ILDM technique, we estimate a local source at the 
transport time-scale and assign it as an error on the QSS assumption 
of CRMs. This error quantifies a deviation from the assumption and 
its influence on the derived quantities of CRM is formulated. 
Furthermore, we propose a novel LC condition based on the error 
with a consistent measure that incorporates the local production.
We apply this new approach to $\text{Ar}$- or $\text{Hg}$-containing 
fluorescent lamps separately. The comparison of the conventional and 
new LC conditions is presented together with the error. In section 
\ref{sec:AtoPla} governing equations are analysed for the plasma 
under investigation and the main points characterising the LC 
particles are discussed. Section \ref{sec:CRM} introduces CRM and 
section \ref{sec:reduction1} shows the analysis in a diagonal basis, 
which is similar to the first step of the ILDM technique. In section 
\ref{sec:projection} we define the error and introduce the novel LC 
condition. Section \ref{sec:setup} describes the $\text{Ar}$ and 
$\text{Hg}$ plasma parameters and section \ref{sec:results} presents 
the results. In section \ref{sec:redchem_sum} summary and conclusion 
are given.

\section{Particle balances}
\label{sec:AtoPla}

The densities of the species are governed by particle balance equations.
For species $i$, it is given by 
\begin{equation}
\frac{\partial n_i}{\partial t} + \vec{\nabla} \cdot \vec{\Gamma_i} = S_i,
\label{eqn:redchempar_bal}
\end{equation}
where $n_i$ is particle density of species, $\vec{\Gamma_i}$ its 
particle flux and $S_i$ its net production rate. The first term 
expresses the time evolution of densities while the second one
represents convective, diffusive transport. The right-hand side 
corresponds to the chemical reactions. This term explicitly couples 
several species to each other. Additionally, the transport term may 
couple them in the case of multicomponent diffusion \cite{Bek12}, 
which is not considered here.

We introduce the notation used in \cite{Mul90/3} in order to compare 
the various chemical processes. Since losses are proportional to 
$n_i$, the source can be written in terms of the destruction and the 
production
\begin{equation}
S_i={\cal P}_i - {\cal D}_i n_i,
\label{eqn:sou_PD}
\end{equation}
where ${\cal{D}}_i$ represents the destruction frequency and 
${\cal P}_i$ is the production rate. In general, these are functions 
of all other densities and various system parameters. In equilibrium, 
the source satisfies
\begin{equation}
S_i=0
\end{equation}
and leads the equilibrium density
\begin{equation}
n_i^{Eqb} = {\cal{P}}_i/{\cal{D}}_i.
\end{equation}
Furthermore, the transport frequency of species $i$ is defined as
\begin{equation}
\nu_{tr,i} = \frac{\vec{\nabla} \cdot \vec{\Gamma}_i}{n_i}.
\end{equation}
As a result, the particle balances can be written as
\begin{equation}
\frac{\partial n_i}{\partial t} + \nu_{tr,i} n_i = {\cal P}_i - {\cal D}_i n_i.  
\label{eqn:PD}
\end{equation}

The ratio of transport to destruction frequencies determines dominant 
process among them. Keeping this in mind, we introduce 
Damk\"{o}hler number $D_{a,i} = \frac{{\cal D}_i}{\nu_{tr,i}}$, then 
equation (\ref{eqn:PD}) further takes the form
\begin{equation}
\frac{1}{{\cal D}_i}\frac{\partial n_i}{\partial t} + \frac{n_i}{D_{a,i}} = 
n_i^{Eqb} - n_i.  
\label{eqn:PDD}
\end{equation}
For large Damk\"{o}hler number, $D_{a,i} \gg 1$ the transport term 
can be ignored in equation (\ref{eqn:PD}):
\begin{equation}
\frac{1}{{\cal D}_i}\frac{\partial n_i}{\partial t} = \frac{{\cal P}_i}{{\cal D}_i} - n_i.
\label{eqn:PD2}
\end{equation}
If ${\cal{P}}_i$ is constant of time, then $n_i^{Eqb}$ is reached 
exponentially in characteristic time $1/{\cal{D}}_i$, which is long 
before the transport time-scale $\tau_{tr}=1/\nu_{tr,i}$. Bates 
{\em et al.} \cite{Bat62} realised that these species, called 
{\em Local Chemistry} (LC) species in CRM, are initially dominated 
by the local source. In this case, the destruction diminishes the 
net production rate and they reach {\em Quasi-Steady State} (QSS) 
\begin{equation}
{S}_i=0. 
\label{eqn:qss}
\end{equation}

In order to manipulate the coupled particle balance equations more 
conveniently, vector notation is introduced. Let 
${\bf n} = \left\{ n_i \right\}$ be the column vector of species 
densities $n_i$. Similarly, the transport and source terms are  
$\vec{\nabla} \cdot \vec{\bf \Gamma}=\left\{ \vec{\nabla} \cdot \vec{\Gamma}_i
\right\}$, ${\bf S} = \left\{ S_i \right\}$. With this notation, 
the particle balance equations can be written as
\begin{equation}
\frac{\partial {\bf n }}{\partial t} + \vec{\nabla} \cdot \vec{\bf \Gamma} = {\bf S}.
\label{eqn:par_bal_vec}
\end{equation}

This work focuses on atomic {\em Electron Excitation Kinetics} (EEK) 
plasmas that are defined by van der Mullen \cite{Mul90/3}. The chemical 
reactions in these plasmas are dominated by radiative and electron 
induced transitions between atomic levels. Besides the ground state, 
the ion and the electron; the excited levels play a role in such 
reactions. Explicit forms of these reactions are listed in table 
(\ref{tab:redchemrea}). One common assumption in EEK plasmas is
that the electrons act as ``external agents'' with density $n_e$ and 
temperature $T_e$ \cite{Mul90/3}. They are not included in density 
vector $\bf n$. 
\begin{table*} 
\begin{tabular}{llll} 
\hline
Label &Reaction & Rate  & Name  \\ \hline \hline
R1 & $A(i) + e \rightarrow A(j) +e, \: (i<j) $   & $n_i n_ek_{ij}$ & Electron impact excitation \\ 
R2 & $A(i) + e \rightarrow A(j) +e, \: (i>j)$   & $n_i n_ek_{ij}$ & Electron impact de-excitation \\ 
R3 & $A(i) + e \rightarrow A^+ +2e $ & $n_in_ek_{i'+'}$ & Electron impact ionisation  \\ 
R4 & $A^+ + 2e \rightarrow A(j) + e $   & $n_{'+'}n^2_eK^{(3)}_{'+'i}$ & Three particle recombination  \\ 
R5 & $A(i) \rightarrow A(j) + h\nu$ & $n_iA^*_{ij}$ & Spontaneous emission  \\ 
R6 & $A(i) + h\nu \rightarrow A(j) + 2h\nu$ & $n_i\beta_{ij}(\nu)$ & Stimulated emission  \\ 
R7 & $A(j) + h\nu \rightarrow A^+ + e$   & $n_{j}\alpha_{j}$ & Photoionisation \\ 
R8 & $A^+ + e \rightarrow A(j) + h\nu $   & $n_{'+'}n_e\alpha_{j}$ & Radiative recombination  \\ \hline \hline
\end{tabular}
\caption{Chemical reactions in atomic pure EEK plasmas. $A$ and $e$ 
represents atom and electrons, while $i$ specify the neutral species 
and $+$ ions. These neutral species are composed of ground and excited 
levels of the atom. In this work, we neglect the photoionisation and 
stimulated emission.} 
\label{tab:redchemrea} 
\end{table*}

Since the source terms of the other species are linear in their 
densities, these can be written as
\begin{equation}
{ \bold S} = {\bold M}(n_e, T_e) {\bold n}.
\label{eqn:source_vector}
\end{equation}
The source matrix $\bold M$ is composed of the reaction frequencies, 
combining equation (\ref{eqn:source_vector}) with table \ref{tab:redchemrea} 
yields
\begin{widetext}
\begin{equation}
M_{ij} = \left\{
\begin{array}{l l l}
 -\left[ \sum_{p \neq j} n_e k_{jp}  + n_e k_{j'+'} \sum_{p \le j} A^*_{jp} \right] &  \quad  
     i=j & \quad \text{R1, R2, R3, R5} \\
 n_e k_{ji} + A^*_{ji} &  \quad  i<j & \quad \text{R1, R5} \\ 
 n_e k_{ji} &  \quad  i>j & \quad \text{R2} \\ 
 n_e k_{ji} &  \quad  i='+' & \quad \text{R3} \\ 
 n^2_e K^{(3)}_{ji} + n_e \alpha_i &  \quad  j='+' & \quad \text{R4, R8} \\ 
\end{array}
\right.
\label{eqn:Mij}
\end{equation}
\end{widetext}
where photoionisation and stimulated emission are not included.  
For $i \neq j$, $M_{ij}$ is the production frequency of state 
$i$ from $j$ and ${M}_{ii}$ is the destruction frequency of state $i$. 
With this notation, the production term in equation (\ref{eqn:sou_PD}) 
is ${\cal P}_i = \sum_{j \neq i}M_{ij} n_j$ while the destruction 
frequency is ${\cal D}_i = M_{ii}$

\section{Collisional radiative models}
\label{sec:CRM}

Solving too many coupled particle balance equations causes heavy 
computational load and stiffness. In order to prevent this, CRMs 
reduce the number of species, for which these equations are solved 
without neglecting the species or their role in the reactions. 
The model relies upon a classification of species into two types: 
{\em Local Chemistry} (LC) and {\em Transport Sensitive} (TS). 
LC species satisfies ${D}_{a,i} \gg 1$ and reaches QSS 
(equation (\ref{eqn:qss})) before the transport time-scale. They 
feature vanishing local sources while the others do not. 

The relations between LC and TS species is re-formulated by van 
Dijk {\em et al.} \cite{Dij2001/1} in the following analysis. 
Reordering the species indices such that the TS species are at the top of 
the vector, ${\bf S}$ and ${\bf n}$ can be decomposed as \cite{Dij2001/1}:
\begin{equation}
{\bf S} =
\left[ \begin{matrix}
 {\bf S}_t \\
 {\bf S}_l
\end{matrix} \right], \: \:
{\bf n} = 
\left[
\begin{matrix}
 {\bf n}_t \\
 {\bf n}_l
\end{matrix}
\right]  ,
\end{equation}
where $l$ labels the LC and $t$ labels the TS species. 
Imposing QSS assumption on the LC species ${\bf S}_l = {\bf 0}$ and
similarly decomposing the source matrix $\bf M$, equation 
(\ref{eqn:source_vector}) becomes
\begin{equation}
\left[ \begin{matrix}
 {\bf S}_t \\
 {\bf 0}
\end{matrix} \right]
=
\left[
\begin{matrix}
  {\bf M}_{tt} & {\bf M}_{lt} \\
  {\bf M}_{tl} & {\bf M}_{ll}
\end{matrix}
\right]
\left[
\begin{matrix}
 {\bf n}_t \\
 {\bf n}_l
\end{matrix}
\right].
\end{equation}
This yields two sets of equations;
\begin{align}
\label{eqn:crm_jan1}
{\bf S}_t = {\bf M}_{tt} {\bf n}_t + {\bf M}_{lt} {\bf n}_{l},\\
{\bf 0} = {\bf M}_{tl} {\bf n}_t + {\bf M}_{ll} {\bf n}_{l}. 
\label{eqn:crm_jan2}
\end{align}
Inverting equation (\ref{eqn:crm_jan2}) shows that the LC densities 
depend linearly on the TS densities 
\begin{equation}
{\bold n_l} = - \left( {\bold M}^{-1}_{ll}  {\bold M}_{tl} \right) {\bold n_t}.  
\label{eqn:crm_lcmap}
\end{equation}
Hence, once the TS densities are known, the complete state of the 
plasma is available with this linear relation. Moreover, putting 
this relation in equation (\ref{eqn:crm_jan1}) we introduce an 
effective TS source term
\begin{equation}
{\bold S}_t = \left( {\bold M}_{tt} - {\bold M}_{lt} {\bold M}^{-1}_{ll}  
{\bold M}_{tl} \right) {\bold n_t}.
\label{eqn:crm_effectivesource}
\end{equation}
The first term on the right-hand side represents the direct 
processes between the TS states, while the second term describes 
{\em ladder-like} transitions, that is via the LC states. Equations 
(\ref{eqn:crm_lcmap}) and (\ref{eqn:crm_effectivesource}) allow to 
take the LC states into account without solving their balance 
equations.

\begin{figure*}[htbp]
\centering
\includegraphics[width=0.70\textwidth]{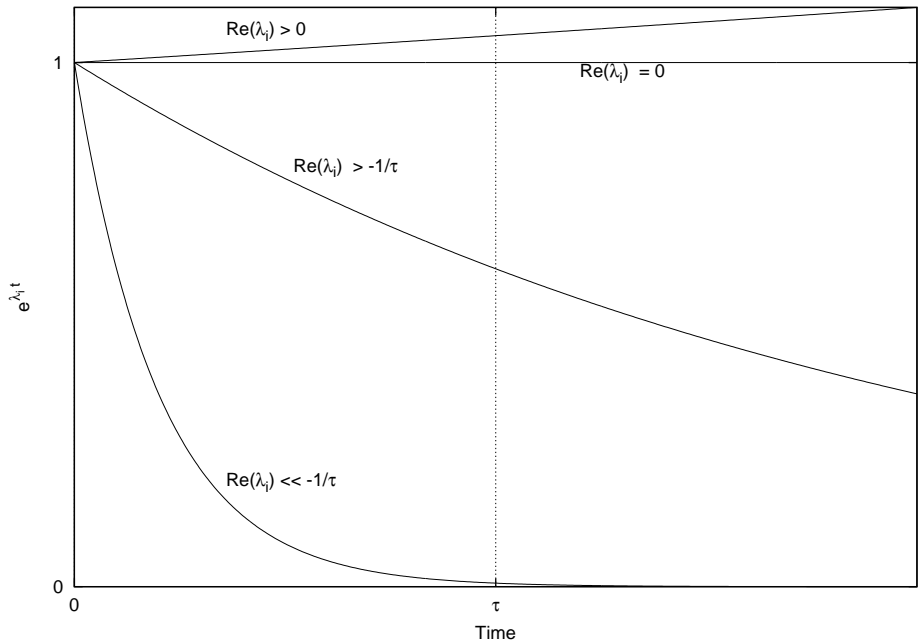}
\caption{
The behaviour of pseudo-density $\bar{n}$ for distinct real 
eigenvalues. The positive eigenvalues cause exponential growth while 
the negative ones induce exponential decay. Comparison with the 
external time-scale $\tau$ reveals pace of the decay. 
}
\label{fig:OrthDensBeh}
\end{figure*}

In spite of their wide usage CRMs contain two weak points. Firstly, 
the classification condition of LC species (${D}_{a,l} \gg 1$) is 
not always sufficient: Since the local production, ${\cal{P}}_l$,
is a function of all other densities 
(see equations (\ref{eqn:PD2}, \ref{eqn:qss})), it is not constant 
in time. As a consequence, the exponential decay to the equilibrium 
density with characteristic time-scale $1/{\cal{D}}_l$ is not valid.
Secondly, the deviations from the assumption ${\bf S}_l={\bf 0}$ are 
not quantified. These deviations may be significant and lead 
perturbations in the LC densities through equation 
(\ref{eqn:crm_lcmap}) that propagate into the TS sources via equation 
(\ref{eqn:crm_effectivesource}). In following sections, we estimate 
these deviations thus define an error for the QSS assumption. 
We also quantify the resultant perturbations on equations 
(\ref{eqn:crm_lcmap}, \ref{eqn:crm_effectivesource}) and propose a 
novel condition to classify the LC species. Rudimentary aspects of 
the estimation and the assignment are developed in section 
\ref{sec:reduction1} while the error and the novel LC condition are 
defined in section \ref{sec:projection}.

\section{Diagonal basis}
\label{sec:reduction1}

Similar to the plasma sciences, the combustion engineering studies 
physical systems containing a large number of species that interact 
with complex chemical kinetics. Aforementioned numerical difficulties 
also arise in the models and Maas {\em et al.} \cite{Maa92} 
introduces ILDM technique to reduce the number of species. Adapting 
the notions of the ILDM technique, we introduce a diagonal basis to 
analyse the particle balances. Settling in this basis, decoupled set 
of equations are acquired with a characteristic time-scale assigned 
to each. This fact allows a classification of those equilibrated at 
a time-scale and forms the backbone of the error definition.

The technique starts with an analysis of source matrix. In order to 
investigate the chemical nature of the particle balances, the 
transport term is omitted in equation ({\ref{eqn:par_bal_vec}}) 
for the time being \cite{Maa92}:
\begin{equation}
\frac{\partial {\bf n}}{\partial t} = {\bf M n}.
\label{eqn:smt}
\end{equation}

\begin{figure*}[t!]
\centering
\includegraphics[width=0.70\textwidth]{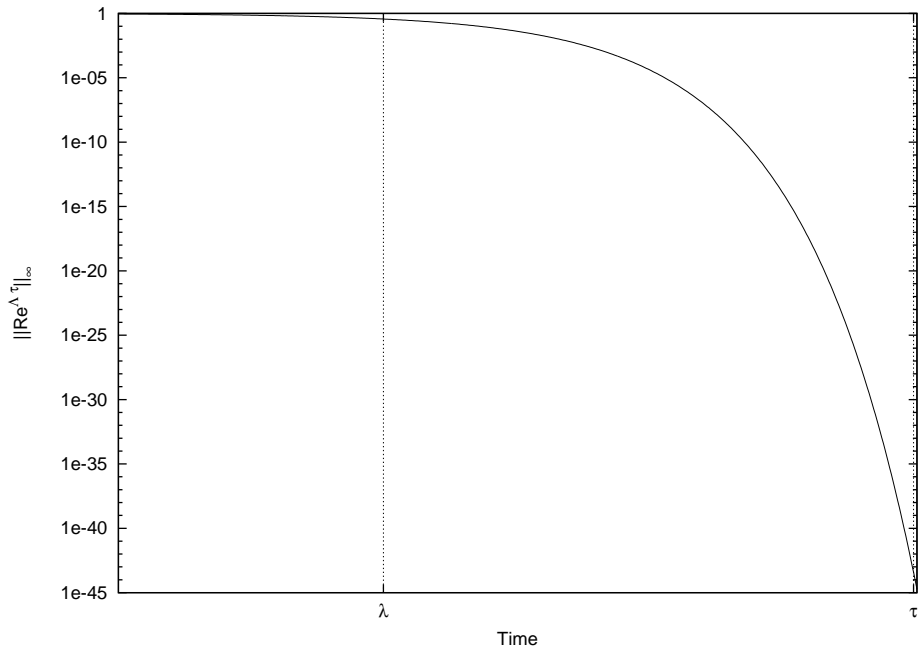}
\caption{
The upper limit for the infinite norm of the fast pseudo-density, 
${\left| \left| \bar{\bf n}_f\right| \right|_{\infty} /n} \leq \left| \left| {\bf R} e^{{\bf \Lambda} \tau}  \right| \right|_{\infty}$, 
through time. For the external time-scale satisfying 
$\left| 1/\lambda \right|=0.01 \tau$, ${\bf n}_f(\tau)$ is negligible.
}
\label{fig:RedErr}
\end{figure*}

Assuming that the matrix ${\bf M}$ is diagonalisable, we have
\begin{equation}
{\bf M}(n_e, T_e) = {\bf V \Lambda V}^{-1}
\label{eqn:dia}
\end{equation} 
where $\bf V$ is a non-singular matrix, whose columns are the eigenvectors, 
and ${\bf \Lambda} = \text{diag}[\lambda_i ]$ is a diagonal matrix 
that contains the corresponding eigenvalues. These matrices are 
functions of $n_e$ and $T_e$ like $\bf M$, although we leave out the 
parameters for readability.

Substitution of equation (\ref{eqn:dia}) into equation (\ref{eqn:smt}) 
and left multiplication with ${\bf V}^{-1}$ yields
\begin{equation}
{\bf V}^{-1} \frac{\partial {\bf n}}{\partial t} ={\bf \Lambda V}^{-1} {\bf n}.
\end{equation}
Furthermore, let $\tau_M$ be the $\bf M$ matrix time-scale, which 
represents how quick it changes, and scales with how fast electron 
density and temperature alters. For time-scales $t \ll \tau_M$, the 
variations of the source matrix are negligible. As a result, its 
eigenvalues and eigenvectors do not vary significantly in time. This 
leads to the form
\begin{equation}
 \frac{\partial \left( {\bf V}^{-1}{\bf n} \right) }{\partial t} ={\bf \Lambda} \left( {\bf V}^{-1} {\bf n} \right).
\label{eqn:ortbal}
\end{equation}
This suggests the introduction of alternative density variables 
$\bar{\bf n}$ according to
\begin{equation}
\bar{\bf n} = {\bf V}^{-1} {\bf n},
\label{eqn:pdd}
\end{equation}
which we call pseudo-densities. Substitution into equation (\ref{eqn:ortbal}) yields
\begin{equation}
\frac{\partial \bar{\bf n}}{\partial t} ={\bf \Lambda}\bar{\bf n}.
\label{eqn:red_eqn}
\end{equation}
Since $\bf \Lambda$ is diagonal, the equations for $\bar{\bf n}$ are 
decoupled and their treatment is relatively easy compared to those 
in original basis. For times $t \ll \tau_M$, the eigenvalues are 
constant and integration of equation (\ref{eqn:red_eqn}) yields 
\begin{equation}
\bar{\bf n}(t) =  e^{{\bf \Lambda} t} \bar{\bf n} (0).
\label{eqn:vec_ort_den}
\end{equation}
In other words, each pseudo-species labelled by $i$ satisfies
\begin{equation}
\bar{n}_i (t) =  e^{\lambda_i t} \bar{n}_i (0),
\label{eqn:ort_den}
\end{equation}
where we used the component notation for the sake of simplicity. 

\subsection{Classification of pseudo-densities}
\label{sec:reduction2}

The particular behaviour of each pseudo-density $\bar{n}_i$ is 
determined by the corresponding eigenvalue $\lambda_i$, the inverse of 
which represents characteristic time-scale of the equation. 
Obviously, a zero eigenvalue yields time invariant behaviour 
$\bar{n}_i(t) = \bar{n}_i(0)$, while a negative real part of the 
eigenvalue imposes exponential decay and estimates a decay frequency, 
$Re|\lambda_i|$ (see Figure \ref{fig:OrthDensBeh}). Comparing the 
corresponding decay time-scale to previously determined time 
parameter $\tau$, the pseudo-densities with smaller time-scales have 
already equilibrated to their asymptotic value: $0$.

Following these observations, we classify three type of pseudo-densities 
for a given time parameter $\tau$, for instance, the 
transport time-scale. (1) Fast pseudo-densities posses eigenvalues 
with \hbox{$Re(\lambda_i) \ll -1/\tau$} and they vanish at the time 
parameter: $\bar{n}_i (\tau) \approx 0$. The approximation 
asymptotically turns into equality. (2) Invariant pseudo-densities 
have a zero eigenvalue, and they are constants in time. (3) Otherwise, 
they are termed as slow pseudo-densities and these are still 
dynamically active.
\begin{equation}
\begin{array}{l l l}
\text{Eigenvalue   } & \text{  Type} & \text{Behaviour} \\ \hline
Re(\lambda_i)  \ll -1/ \tau & \text{ Fast } & \bar{n}_i (\tau) \approx 0 \\ 
\lambda_i  =  0 & \text{ Invariant } &  \frac{ d \bar{n}_i}{dt} =0 \\
\text{Else} & \text{ Slow } & \text{Not known}. 
\label{eqn:redchemreduction}
\end{array}
\end{equation}

\begin{figure*}[t!]
\centering
\includegraphics[width=0.70\textwidth]{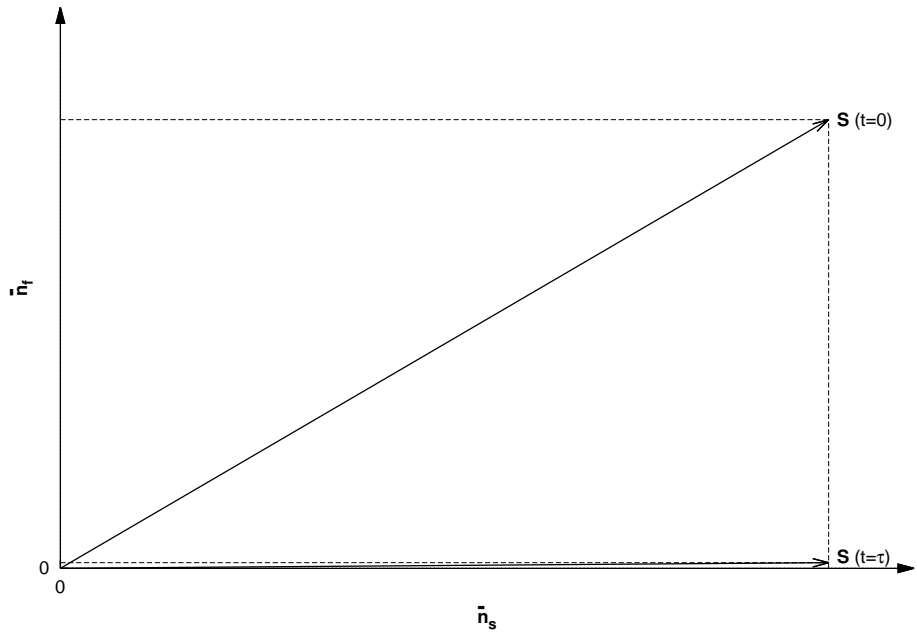}
\caption{A depiction of the source $\bf{S}$ in the density space that 
is spanned by pseudo-densities. The fast component vanishes at 
time-scale $\tau$.
}
\label{fig:Geo}
\end{figure*}

The pseudo-densities are linear combinations of the real densities
and follow certain characteristics of the source matrix. 
The time invariant pseudo-densities represent the conservation of 
the total number of the particles in the (sub-) set of reactions. 
Decaying pseudo-densities represent the equilibrium of chemical 
processes. The imaginary part of the eigenvalue means oscillatory 
pseudo-density while the positive real part induces exponential 
growth. The oscillatory and growing pseudo-densities are not 
observed for the analysed systems, hence they are not treated. 
Since most of the physical systems eventually equilibrate, the 
absence of these eigenvalues is expected.

\subsection{Reconfiguration in the density space}
\label{sec:reduction3}

In this section, we reconfigure the density space based on the 
classification of the pseudo-densities at the time-scale $\tau$.
Firstly, we introduce a diagonal reduction matrix ${\bf R}(\tau)$
\begin{equation}
R_{ii} = \left\{
\begin{array}{l l}
1 & \quad \text{ if } i \in \left\{ \text{ ``Fast''} \right\}  \\
0 & \quad \text{ else}.
\end{array}
\right.
\end{equation}
This matrix defines fast $\bar{\bf n}_f (\tau)$ and slow $\bar{\bf n}_s (\tau)$ 
components of the pseudo-density $\bar{\bf n} (\tau)$: 
\begin{align}
\label{eqn:nf}
\bar{\bf n}_f (\tau) & = {\bf R}(\tau) e^{{\bf \Lambda} \tau} \bar{\bf n} (0), \\
\bar{\bf n}_s (\tau) & = ({\bf I} -{\bf R}(\tau)) e^{{\bf \Lambda} \tau} \bar{\bf n} (0),
\label{eqn:ns}
\end{align}
where $I$ is the identity matrix and $\bar{\bf n}$ satisfies
\begin{equation}
\bar{\bf n} (\tau) = \bar{\bf n}_s (\tau) + \bar{\bf n}_f (\tau).
\label{eqn:project1}
\end{equation}
The fast component $\bar{\bf n}_f (\tau)$ contains all fast pseudo-densities 
and zero otherwise, while $\bar{\bf n}_s (\tau)$ includes 
the rest. Furthermore, they associate with the original density by 
the relations
\begin{align}
\label{eqn:nfn}
\bar{\bf n}_f (\tau) & = {\bf R}(\tau) {\bf V}^{-1} {\bf n} (\tau), \\
\bar{\bf n}_s (\tau) & = \left( {\bf I} -{\bf R}(\tau) \right) {\bf V}^{-1}  {\bf n} (\tau).
\label{eqn:nsn}
\end{align}

This form partitions any density-dependent quantity into the fast and 
the slow components. Accordingly, using the relation 
(\ref{eqn:source_vector}), the source at $\tau$ is given by 
\begin{equation}
{\bf S} (\tau) = {\bf M V}  \bar{\bf n}_s(\tau) + {\bf M V} \bar{\bf n}_f(\tau).
\end{equation}

This configuration is based on the two different regions of the
density space, and each one is described by one type of the pseudo-density. 
(1) The fast region is emptied before $\tau$ and, for any
density-dependent quantity, its corresponding component is depleted. 
(2) The slow one, however, is dominantly occupied around the 
time-scale $\tau$. 

The fast pseudo-densities are classified by their vanishing value at 
$\tau$ (see Figure \ref{fig:RedErr}) and imposing this, the pseudo-density 
$\bar{\bf n}(\tau)$ is approximately described by the slow 
component 
\begin{equation}
\bar{\bf n} (\tau) \approx \bar{\bf n}_s (\tau).
\label{eqn:project2}
\end{equation}
Similarly, implementing this in the source components leads to the 
approximation (depicted in Figure \ref{fig:Geo})
\begin{equation}
{\bf S} (\tau) \approx {\bf M V} \bar{\bf n}_s (\tau).
\end{equation}
Furthermore, we express the source in terms of the original density,
by defining a slow matrix,  
${\bf M^s} (\tau)= {\bf M V} \left( {\bf I}-{\bf R} \right) {\bf V}^{-1}$,
\begin{equation}
{\bf S} (\tau) \approx  {\bf M^s n} (\tau). 
\label{eqn:slow_source2}
\end{equation}

\subsection{Role of the transport} 
\label{sec:reduction5}

In the previous analysis, we omitted the transport in the particle 
balances and focused on the local source. In this section, we 
introduce an effective transport term and analyse its effect on 
the pseudo-densities. Firstly, we assign a hypothetical transport 
frequency to all species 
\begin{equation}
\vec{\bf \nabla} \cdot \vec{\bf \Gamma} = \nu_{tr} {\bf n},
\end{equation}
and the particle balances take the form
\begin{equation}
\frac{\partial  {\bf n}}{\partial t} = {\bf M n} -\nu_{tr} {\bf n}.
\end{equation}
By diagonalisation of $\bf M$ we acquire equation 
(\ref{eqn:ort_den}) in a new form 
\begin{equation}
\bar{n}_i (\tau) =  e^{\left( \tau \lambda_i - \tau \nu_{tr} \right) } \bar{n}_i (0),
\label{eqn:ort_den_tr}
\end{equation}
which defines the role of the effective transport on the pseudo-densities.

Since the fast pseudo-species are defined by the relation 
$\tau \left| Re(\lambda_i)\right| \gg 1$, the transport frequency 
that satisfies $\tau \nu_{tr} \leq 1$ does not affect their behaviour.
If it fulfils $\tau \nu_{tr} \gg 1$, the transport is relatively quick 
and the fast pseudo-density vanished long before time $\tau$. 
Otherwise, the transport does not play a role. Furthermore, such a 
transport imposes that the invariant pseudo-density is no longer 
constant in time but exponentially decreases. In the case that 
$\tau \nu_{tr} \gg 1 $, it is classified as a fast pseudo-density and 
otherwise it is a slow pseudo-density. A transport frequency such 
that $\tau \nu_{tr} > 1$ switches the slow pseudo-densities to fast 
if it also satisfies 
$\tau  \gg 1/(\left| Re(\lambda_i)\right| + \nu_{tr})$.
On the other hand, the slow behaviour does not change for 
$\tau  \nu_{tr} \leq 1$, unless it is very close to being classified as 
fast. 

In conclusion, the transport does not change the behaviour and
classification of the pseudo-densities for $\tau / \nu_{tr} \leq 1$.
Otherwise, it may shift the fast behaviour earlier in time and should 
also be taken into account. This implies that the source approximation 
is still valid at $\tau = \tau_{tr}$, independent of the transport
\begin{equation}
 {\bf S } {(\tau_{tr})}  \approx
({\bf M^s n})(\tau_{tr}). 
\end{equation}
As a result, we have an approximate density-dependent source that is 
defined at the transport time-scale $\tau_{tr}$.

Note that the transport frequency certainly differs for ions due to 
the ambipolar electric field. Since the ion transport frequency is 
larger compared to that of neutrals \cite{Rog85}, 
$\tau / \tau_{tr,ion} < 1$ is already satisfied if 
$\tau / \tau_{tr} \leq 1$.

\begin{figure*}[t!]
\centering
\includegraphics[width=0.80\textwidth]{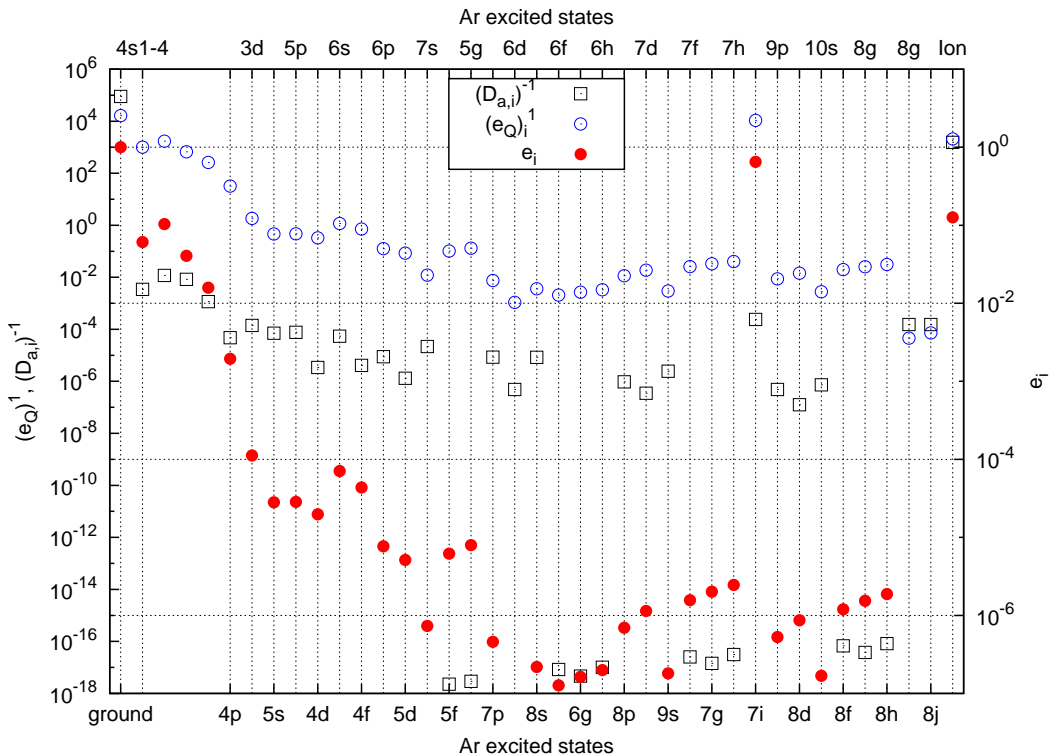}
\caption{
(Argon plasma) The density-independent consistent QSS error $({\bf e_Q})_i^1$, the inverse of 
Damk\"{o}hler number $(D_{a,i})^{-1}$ and the error number $e_i$ for 
numbered species $(i)$. Note the double y-axes with different scales.
The transport time-scale is $\tau_{tr} = 5.20 \times 10^{-04} \: $s 
calculated from equation (\ref{eqn:TrFreq}).
}
\label{fig:ArAve}
\end{figure*}

\section{QSS Error of LC levels}
\label{sec:projection}

CRMs assume that the LC levels have already reached QSS at 
$\tau_{tr}$, due to their large Damk\"{o}hler number: 
\begin{equation}
\left. {\bf S_{l} }\right|_{\tau_{tr}}  = 
{\bf 0}, 
\end{equation}
which is based on a production term that is constant in time
(see equations (\ref{eqn:PD2}, \ref{eqn:qss})). Furthermore, the 
analysis in the diagonal basis assigns an approximate, 
density-dependent nonzero value to the source
\begin{equation}
\left. {\bf S_{l} }\right|_{\tau_{tr}}  \approx
({\bf M^s n})_l(\tau_{tr}). 
\end{equation}
Regarding this, we use the latter to quantify the deviation from the 
assumption and define its absolute as the error that is made by the 
QSS assumption of the LC levels. Consistent QSS error at the transport 
time-scale is given by the relation
\begin{equation}
 ({\bf e_Q})_l (\tau_{\text{tr}}) = \left| ({\bf M^s n})_l \right|,
\end{equation}
where $|.|$ represents the absolute value.

Furthermore, replacing it with zero sources in equation 
(\ref{eqn:crm_jan1}), the error also propagates through the following 
CRM relations. The mapping between TS and LC 
(equation (\ref{eqn:crm_lcmap}))  perturbs with 
\begin{equation}
{\bf \delta n}_l (\tau_{\text{tr}}) = {\bf M}_{ll}^{-1} ({\bf e_Q})_l
\end{equation} 
while the effective TS source - equation 
(\ref{eqn:crm_effectivesource}) - perturbs with
\begin{equation}
{\bf \delta S}_t (\tau_{\text{tr}}) = {\bf M}_{lt} {\bf M}^{-1}_{ll} ({\bf e_Q})_l.
\end{equation}

\subsection{Density-independent QSS error and novel LC condition}

Let $\bf M^s_i$ be the $i_{th}$ row of $\bf M^s$. The QSS error, 
$({\bf e_Q})_i$, is the inner product of $\bf M^s_i$ and $\bf n$ and 
it satisfies the H\"{o}lder's inequality
\begin{equation}
({\bf e_Q})_i  (\tau_{tr}) \leq \left| \left| {\bf M^s_i} \right| \right|_1 \left| \left| {\bf n} \right| \right|_1,
\end{equation}
where $||.||_1$ represents the 1-norm of a vector. Since the 1-norm 
of the density vector is the total density, $n$, this reduces to the 
form 
\begin{equation}
({\bf e_Q})_i (\tau_{tr}) \leq ({\bf e_Q^1})_i(\tau_{tr}) n,
\label{eqn:hol}
\end{equation}
where we define density-independent consistent QSS error of LC 
species $i$ 
\begin{equation}
 ({\bf e_Q^1})_i(\tau_{tr}) = \left| \left| {\bf M^s_i}(\tau_{tr}) \right| \right|_1.
\end{equation}
Together with $n$, it determines the upper bound for the 
density-dependent error and in case the density values are not available, 
it can be used instead.

Furthermore, we define a novel LC condition based on the 
density-independent consistent QSS error. In order to overcome the arbitrary 
scale of the error, which is intrinsically determined by each system,
a dimensionless QSS error is defined by the relation
\begin{equation}
\widehat{({\bf e_Q})}_i (\tau_{tr}) = \frac{({\bf e_Q})_i(\tau_{tr})}{n \zeta},
\end{equation}
where $\zeta$ is a non-dimensionalisation constant with unit $1/$s. 
The inequality (\ref{eqn:hol}) suggests that
\begin{equation}
\widehat{({\bf e_Q})}_i(\tau_{tr}) \leq e_i(\tau_{tr}),
\end{equation}
where we define a dimensionless error number
\begin{equation}
e_i (\tau_{tr})= \frac{({\bf e_Q^1})_i (\tau_{tr})}{\zeta}.
\end{equation}
If this number is sufficiently small compared to unity, it ensures a 
negligible dimensionless QSS error relative to $\zeta$. In this 
respect, we classify LC level as a species that satisfy 
\begin{equation*}
\text{If } \: e_i \ll 1 \: \: \text{ then } i \text{ is an LC level. }
\end{equation*}
For the $\zeta$, we use the density-independent error of the ground 
state, $({\bf e_Q}^1)_0$, with the assumption that the ground state 
is a TS level with the highest density-independent error. Since the 
ions do have complex transport phenomena, such as ambipolar diffusion,
we do not classify them but assume that they are TS.

\begin{figure*}[t!] 
\centering
\includegraphics[width=0.80\textwidth]{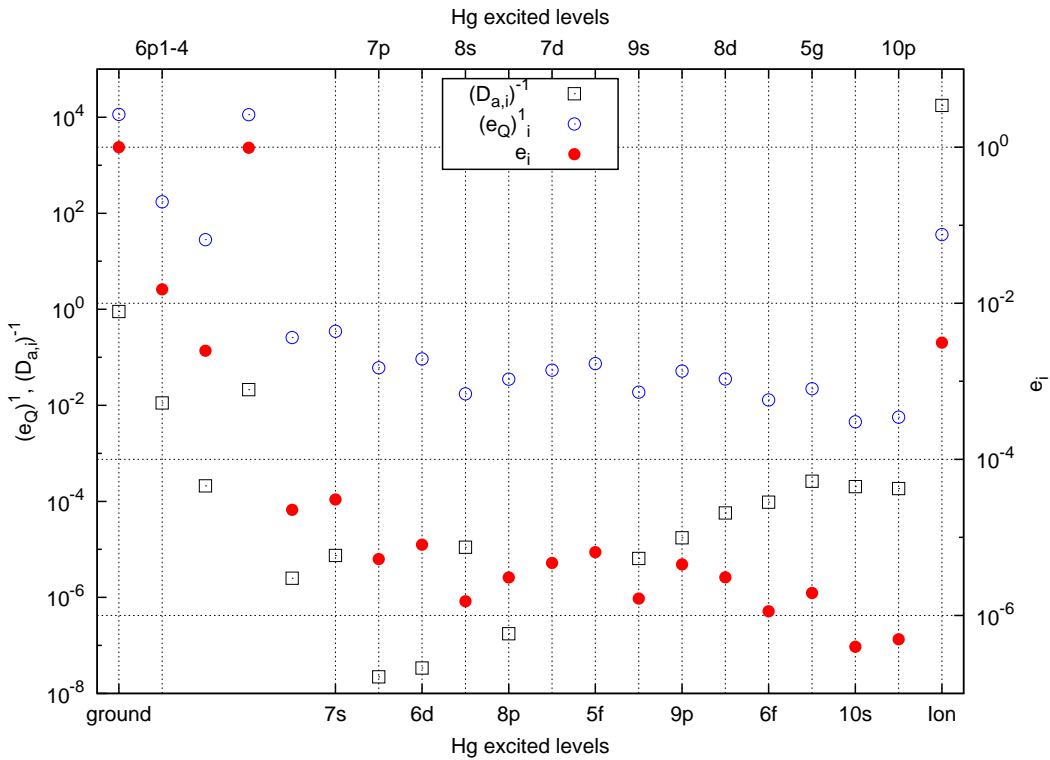}
\caption{
(Mercury plasma) The density-independent consistent QSS error $({\bf e_Q})_i^1$, the inverse of 
Damk\"{o}hler number $(D_{a,i})^{-1}$ and the error number $e_i$ for 
numbered species $(i)$. Note the double scales. Transport time-scale 
is $\tau_{tr} = 1.17 \times 10^{-03} \: $s, which is determined via 
equation (\ref{eqn:TrFreq}). 
\label{fig:HgAve}
}
\end{figure*}

\begin{table}
\begin{center}
\begin{tabular}{ccc}
\hline
Parameters & $\text{Ar}$ & $\text{Hg}$  \\ \hline \hline
$\sigma$   & $ 10^{-19} \: \text{m}^{2} $ &  $ 10^{-19} \: \text{m}^{2}$ \\
$n$ & $ 10^{22} \: \text{m}^{-3} $ & $ 10^{22} \: \text{m}^{-3} $  \\ 
$R$ & $ 2.0 \times 10^{-2} \: \text{m} $ & $ 2.0 \times 10^{-2} \: \text{m} $ \\ 
$L$   & $1.80 \: \text{m}$ & $1.80 \: \text{m}$  \\ 
$T_e$   & $1 \: \text{eV}$ & $1 \: \text{eV}$  \\ 
$T_h$   & $300 \: \text{K}$ & $300 \: \text{K}$  \\ 
$\tau_{tr}$ & $ 5.20 \times 10^{-04} \: \text{s} $ & $ 1.17 \times 10^{-03} \: \text{s} $ \\ 
$n_{e}$ & $ 1.0 \times 10^{18} \: \text{m}^{-3} $ & $ 1.0 \times 10^{18} \: \text{m}^{-3} $ \\ 
\hline\hline 
\end{tabular}
\label{tab:pla_para}
\caption{Main EEK atomic plasma parameters.}
\end{center}
\end{table}

\section{Model setup} 
\label{sec:setup} 

The QSS errors and the error numbers are investigated in the 
cylindrical column of a fluorescent lamp. The plasma is contained in 
the radius $R=2.0 \times 10^{-2} \: $m within the length 
$L=1.80 \: $m. We analyse either argon or mercury based atomic 
discharges, neglecting spatial inhomogeneities and the sheath 
phenomena. Additionally, we assume the two-temperature plasma 
condition of the Maxwellian distribution functions, together with the 
electron temperature $T_e = 1 \: $eV, the gas temperature 
$T=300 \: $K and the electron density 
$n_e=1.0 \times 10^{18} \: \text{m}^{-3}$.  

The atomic excited states above certain threshold energy are neglected 
in the model. In this way, an infinite number of species is reduced to 
a reasonable finite set. For the discussion and the methods for these 
neglected levels, we refer to a study by van Dijk {\em et al.} 
\cite{Dij2000/1}. The $\text{Ar}$ system contains totally $39$ species 
including the ground, the excited states and the ion. The excited 
levels are chosen from smallest energy $\text{Ar}(4s)$ group up to the
energy level $\text{Ar}(8j)$. Transition rates between these species 
are taken from \cite{Ben93}. Together with the ground state and the 
ion, $\text{Hg}$ plasma contains $20$ species. The excited levels 
range from $\text{Hg}(6p1)$ group with the lowest energy to the level 
$\text{Hg}(10p)$. The corresponding transition rates are adapted from 
\cite{Dij2001/2}.

\subsection{Transport frequency}

The transport frequency is determined from the flux term in the 
particle balances. A fluorescent lamp has negligible convective flux 
and the diffusive transport satisfies \cite{Cha87} 
\begin{equation}
\left|\vec{\nabla} \cdot D \vec{\nabla} n_i \right| = \frac{D}{\Lambda^2} n_i,
\label{eqn:TrFreq}
\end{equation} 
where $\Lambda$ is the diffusion length and $D$ is the diffusion 
coefficient. The coefficient $D$ relates with the mean free path of 
elastic collisions $\lambda$ and thermal velocity $v_{th}$:
$D=1/3 \lambda v_{th}$. The mean free path is 
$\lambda =\frac{1}{n \sigma}$, where $n$ is the dominant gas density 
and $\sigma$ is the momentum transfer cross-section. The thermal 
velocity satisfies $v_{th}=\sqrt{8k_BT_h/\pi m}$ where $T_h$ is the 
gas temperature, $m$ is the particle mass and $k_B$ is the Boltzmann 
constant. We assume that the fundamental diffusion mode dominates 
\cite{Cha87}:
\begin{equation}
\frac{1}{\Lambda^2} = \frac{1}{\Lambda_0^2} =
\left( \frac{\pi}{L}\right)^2 + \left( \frac{2.405}{R}\right)^2
\end{equation} 
because the mean free path is smaller regarding the container
dimensions. As a result, the transport frequency of neutrals is given 
by the relation 
\begin{equation}
\nu_{tr} =  \frac{\sqrt{8k_BT_h/\pi m}}{3 n \sigma \Lambda_0^2}.
\end{equation}

\begin{figure*}[t!]
\centering
\includegraphics[width=0.80\textwidth]{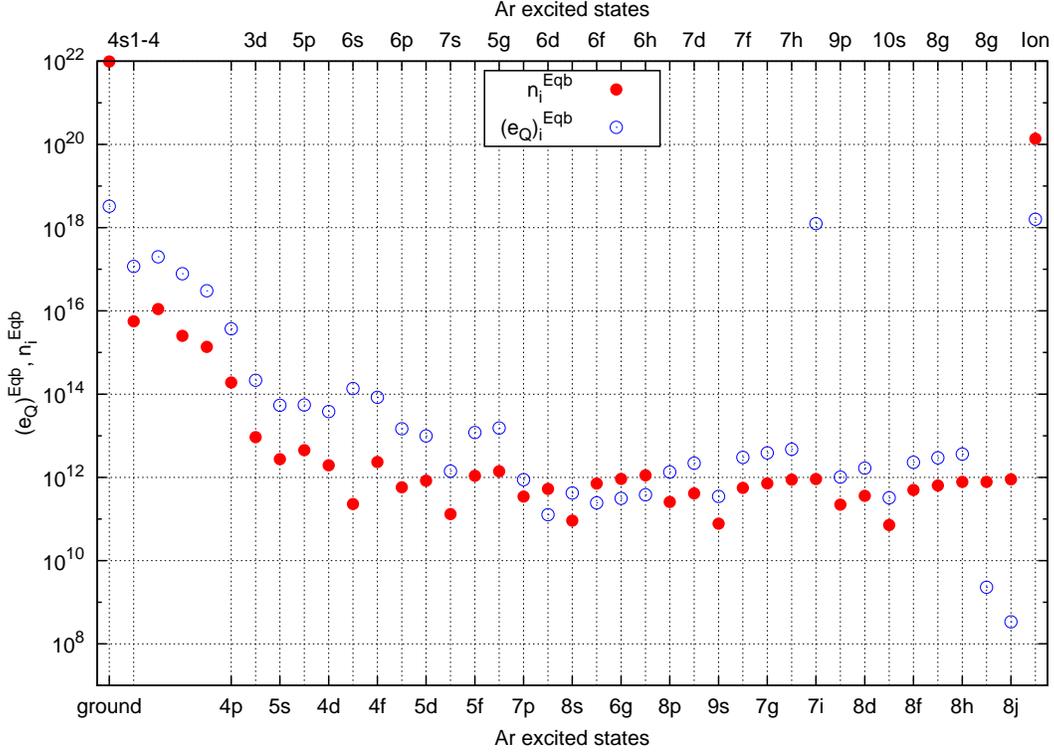}
\caption{
The consistent QSS error $({\bf e_Q})_i^{Eqb}$ for equilibrium density 
$n_i^{Eqb}$ of the argon plasma. The transport time-scale is 
$\tau_{tr} = 5.20 \times 10^{-04} \: $s.
}
\label{fig:ArEqb}
\end{figure*}

\section{Results}
\label{sec:results}

We analyse the positive column of a fluorescent lamp for $\text{Ar}$ 
and $\text{Hg}$ respectively. In the analysis, we define that a 
parameter $r \ll 1$ if $r \leq \gamma$, where we set 
$\gamma = 1.0 \times 10^{-2}$. Within this definition, the density-independent 
QSS errors and the error numbers as well as the inverse of the 
Damk\"{o}hler numbers are presented at the calculated transport 
frequencies. (The inverse values are shown for the sake of simplicity.)
The classification of the species in CRM formalism is provided
while the conventional and the novel conditions are compared. The 
maximum scale of the consistent QSS errors are presented for the 
{LC} levels and the consistent QSS errors are also shown 
for equilibrium density values.

The density-independent consistent QSS errors $({\bf e_Q})^1_i$, the error 
numbers $e_i$ and the inverse of Damk\"{o}hler numbers $(D_{a,i})^{-1}$ 
for $\text{Ar}$ plasma are presented in Figure \ref{fig:ArAve}. The 
transport time-scale is calculated as 
$\tau_{tr}=5.20 \times 10^{-04} \: $s, using equation 
(\ref{eqn:TrFreq}). In the system, the $\text{Ar}(4s)$ group satisfies 
the QSS requirements according to the conventional condition. However, 
they have larger error numbers $e_i> 1.0 \times 10^{-2}$, hence, 
the novel condition suggests that they are TS instead. The rest of the 
species are classified LC, in agreement with the conventional method. 
Though $\text{Ar}(7i)$ is defined as an LC level, it does not satisfy 
necessary conditions in both methods. We suspect that it is caused by 
a poor formulation of the regarding transitions. Within LC species, 
$({\bf e_Q})^1_i$ determines a maximum value of QSS error 
$1.0 \times 10^{18} \text{m}^{-3}\text{s}^{-1}$ for 
$\text{Ar}(8g), \text{Ar}(8j)$ and between 
$1.0 \times 10^{20} \text{m}^{-3}\text{s}^{-1}$ and
$1.0 \times 10^{24} \text{m}^{-3}\text{s}^{-1}$ for the others. 

Same quantities for $\text{Hg}$ plasma are plotted at transport
time-scale $\tau_{tr}=1.0 \times 10^{-03} \: $s in Figure 
\ref{fig:HgAve}. In this case, both conditions are in agreement. All 
excited levels qualify to be LC, while $\text{Hg}(6p1)$, 
$\text{Hg}(6p3)$ are TS levels. Upper bound of the density-dependent 
error ranges between 
$1.0 \times 10^{20} \: \text{m}^{-3}\text{s}^{-1}$ 
and $1.0 \times 10^{23} \: \text{m}^{-3}\text{s}^{-1}$, where large 
values belong to those excited levels neighbouring ground state.

In both systems, errors and error numbers decrease with the energy of 
the level in average while its slope lessens with energy. $D_{a,i}$ 
also follows a similar trend, while it shows colossal jumps at various 
levels compared to those of errors. We think that this behaviour is 
an effect of the local production comparable to the destruction at the specified 
times-scale. Both ground states have the highest error and numbers.
The ions, on the other hand, are not classified by the number but
assumed to be TS due to its complex transport properties.

\begin{figure*}[t!]
\centering
\includegraphics[width=0.80\textwidth]{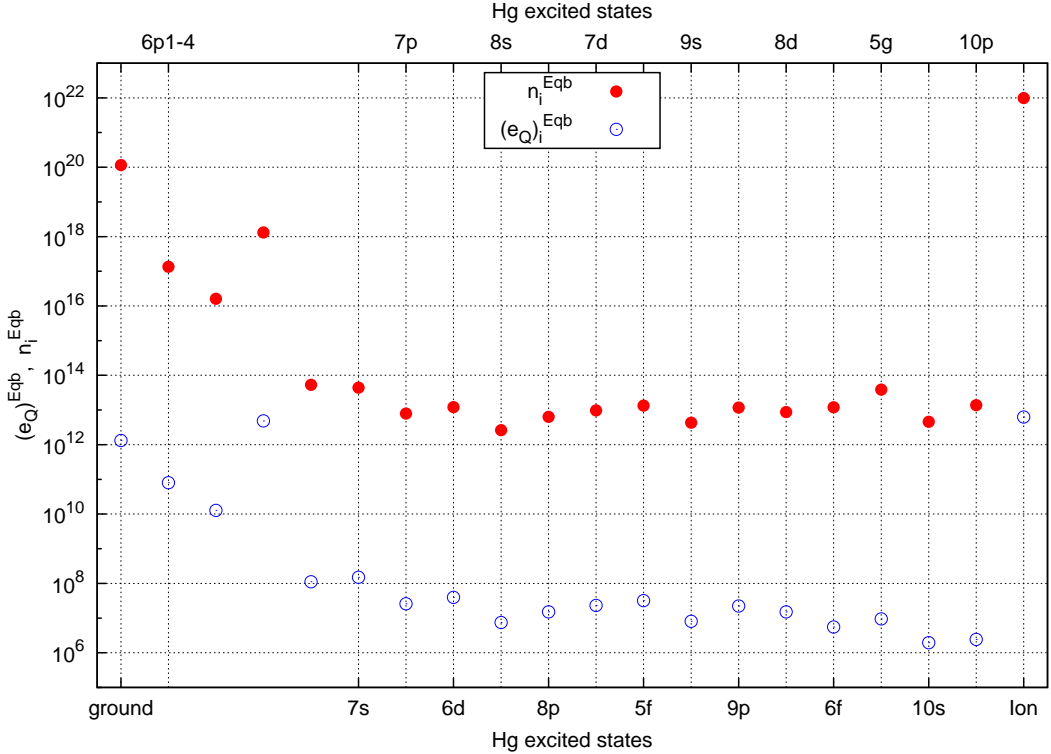}
\caption{
The consistent QSS error $({\bf e_Q})_i^{Eqb}$ for equilibrium 
densities $n_i^{Eqb}$ of the mercury plasma. The transport time-scale 
is $\tau_{tr} = 1.17 \times 10^{-03} \: $s.
}
\label{fig:HgEqb}
\end{figure*}

We also present the consistent QSS errors $({\bf e_Q})_i^{Eqb}$ at the 
equilibrium density distributions $n_i^{Eqb}$ in Figures 
\ref{fig:ArEqb} and \ref{fig:HgEqb}. The density distribution 
reduces the error compared to maximum values given by 
$({\bf e_Q})_i^{1}$. The decrement is with a factor of 
$1.0 \times 10^{-6}$ for argon while it is about 
$1.0 \times 10^{-10}$ for mercury. On the other hand, their values 
relative to each other does not change.

\section{Summary and conclusion}
\label{sec:redchem_sum}

We analyse atomic EEK plasmas and focus on the LC levels of the CRMs.
These levels are assumed to be in QSS at a transport time-scale due 
to their large Damk\"{o}hler number $D_{a,i} \gg 1$. This assumption 
is only valid in the case of a time-independent local production. However, 
the production implicitly depends on time via the densities. 
Furthermore, in CRMs, any kind of deviation from exact QSS behaviour 
and its role on the rest of the system is not quantified. In order to 
tackle this problem and describe the deviation, we analyse the plasma 
particle balance equations of all species in the diagonal basis of 
their source Jacobian. This analysis provides an approximate 
density-dependent source at the transport time-scale and we set its absolute 
value as the error of QSS assumption on the LC levels. In case the 
densities are not foreknown, we further define a density-independent error, 
which forms an upper bound of the error. Additionally, a 
dimensionless number is provided and sufficiently small number
minimises the error and identifies a novel LC condition.

The technique is applied to the positive column of a fluorescent lamp with 
$\text{Ar}$ and $\text{Hg}$ separately. For the defined transport 
time-scale, $\text{Ar}(4s)$ group is identified to be a TS level, 
in contrast with the conventional condition. The rest possesses low error 
number $e_i < 0.01$, which classify them as LC levels.
$\text{Ar}(7i)$ substantially deviates from this. Though solid 
evidence lacks, it is reasonable that poor transition rate causes 
this deviation. The density-dependent error maxima ranges between 
$7.0 \times 10^{18} \: \text{m}^{-3}\text{s}^{-1}$ and
$1.0 \times 10^{24} \: \text{m}^{-3}\text{s}^{-1}$.
For $\text{Hg}$ system, all excited states are LC species, 
except $\text{Hg}(6p1)$, $\text{Hg}(6p3)$ levels. The error maxima, in 
this case, ranges from
$1.0 \times 10^{20} \: \text{m}^{-3}\text{s}^{-1}$ to
$1.0 \times 10^{23} \: \text{m}^{-3}\text{s}^{-1}$.

In both systems, the density-independent error and the number decline 
with the energy of the excited level. This is in agreement with the 
fact that the energetic levels quickly settle in QSS compared to those 
with less energy. Additionally, the conventional QSS condition works 
for most of the levels but it can be inefficient for the low energy 
levels. For more accurate classification, the novel condition should 
be preferred.

The consistent QSS errors $({\bf e_Q})_i^{Eqb}$ are also shown for
equilibrium density distribution $n_i^{Eqb}$. Compared to the upper 
bound $({\bf e_Q})_i^{1}$, their scale is significantly smaller. In 
this respect, the density-dependent errors should be used whenever the 
densities are available. On the other hand, the distribution 
$n_i (\tau_{tr})=n_i^{Eqb}$ should be used with care, since it is 
only valid in the complete equilibrium.

We defined a $\gamma$ parameter to describe a number that is 
negligibly small compared to unity. In this work, we set it to 
$\gamma= 1.0 \times 10^{-2}$, yet its value is arbitrary. If smaller 
$\gamma$ is chosen, then the description of a negligibly small value 
is further ensured. Additionally, it can set all species TS with a 
cost of higher computational burden in the CRM. Higher $\gamma$ values 
can classify more excited species into LC levels, but they may still 
have large density-independent errors. In this case, the novel 
condition can exactly proceed parallel to the conventional technique.
If it is too high (above $0.1$) even the fast pseudo-species may not 
be negligible. At this point, the choice is left to the user. 
Depending on their needs, $\gamma$ can be kept higher, while the 
resultant effect is defined by the QSS error. We suggest that this 
value should not exceed $0.1$ since the source approximation is 
compromised by larger fast pseudo-densities.

The error is not determined continuously but rather discretely with 
the transport time-scale. The distribution of the eigenvalues 
determines these discrete points that should be also taken into 
account to compute the errors. In the analysed systems, the 
classification of the pseudo-densities and the errors are valid 
between
$ 1.0 \times 10^{-5}\:\text{s} < \tau_{tr} < 5.0 \times 10^{-3}\:$s 
for $\text{Ar}$ and 
$ 1.0 \times 10^{-3}\:\text{s}  < \tau_{tr} < 5.0 \times 10^{-2}\:$s 
for $\text{Hg}$. As a caveat, around these discrete points the 
classification may change due to the arbitrary $\gamma$ parameter. 
Furthermore, in this study the role of the initial density 
configuration is not taken into account in the classification. 
Excessively low initial pseudo-densities may qualify to be fast even 
before the criteria is met. Hence, the error and the error number can 
be misleading around the discrete points of the eigenvalue 
distribution.

In the density-independent definitions, we choose the smallest 1-norm 
among the p-norms. This ensures that the maxima are kept closer to 
the density-dependent values compared to higher norms. Furthermore, 
the non-dimensionalisation constant $\zeta$, which defines the error 
number $e_i$ and the novel LC condition, imposes a QSS definition that 
is relative to $\zeta$. In this study, $\zeta$ is chosen to be the 
maximum dimensionless QSS error. We observe that a different $\zeta$ 
parameter, such as the transport frequency of the plasma, increases the error 
number and changes the classification of the low energy states 
Ar$(4p)$ and Hg$(6p2)$. 

The system is studied for reactions, where two distinct types of 
species interact. Otherwise, a source term that is a nonlinear function 
of the density vector appears. In the presence of these reactions, 
the method can be treated with source term linearisation within 
iterative schemes \cite{Pat80, Egg95}.

\bibliographystyle{unsrt}
\bibliography{refredchem}

\end{document}